\documentclass[%
reprint,
 bibnotes,
 amsmath,amssymb,
 aps,
 prl
]{revtex4-1}

\newcommand*{\ov}{$\mathrm{V_{O}^{\bullet\bullet}}$}
\newcommand*{\cuo}{CuO$_{2}$}
\newcommand*{\degC}{$\!^\circ$C}
\newcommand*{\wn}{$\mathrm{cm^{-1}}$}
\newcommand*{\tc}{T$_{\mathrm{c}}$}
\newcommand*{\aog}{$A_{1g}$}
\newcommand*{\asg}{$A^{*}_{1g}$}

\usepackage{graphicx}
\usepackage{dcolumn}
\usepackage{bm}

\begin{document}


\title{Selective formation of apical oxygen vacancies in La$_{2-x}$Sr$_{x}$CuO$_{4}$}
	
\author{Gideok Kim}
 \email{G.Kim@fkf.mpg.de}
 \affiliation{Max-Planck-Institute for Solid State Research, Heisenbergstrasse 1, 70569 Stuttgart, Germany}%
\author{Georg Christiani}%
 \affiliation{Max-Planck-Institute for Solid State Research, Heisenbergstrasse 1, 70569 Stuttgart, Germany}%
\author{Gennady Logvenov}%
 \email{G.Logvenov@fkf.mpg.de}
 \affiliation{Max-Planck-Institute for Solid State Research, Heisenbergstrasse 1, 70569 Stuttgart, Germany}%
\author{Sungkyun Choi}%
 \affiliation{Max-Planck-Institute for Solid State Research, Heisenbergstrasse 1, 70569 Stuttgart, Germany}%
\author{Hun-Ho Kim}%
 \affiliation{Max-Planck-Institute for Solid State Research, Heisenbergstrasse 1, 70569 Stuttgart, Germany}%
\author{M. Minola}%
 \affiliation{Max-Planck-Institute for Solid State Research, Heisenbergstrasse 1, 70569 Stuttgart, Germany}%
\author{B. Keimer}%
 \affiliation{Max-Planck-Institute for Solid State Research, Heisenbergstrasse 1, 70569 Stuttgart, Germany}%

\date{\today}

\begin{abstract}
The superconducting properties of high-\tc\ materials are functions of carriers concentration, which is controlled by the concentration of defects including heterovalent cations, interstitial oxygen ions, and oxygen vacancies. 
Here we combine low-temperature thermal treatment of La$_{2-x}$Sr$_{x}$CuO$_{4}$ epitaxial thin films and confocal Raman spectroscopy to control and investigate oxygen vacancies. 
We demonstrate that the apical site is the most favorable position to accommodate oxygen vacancies under low-temperature annealing conditions.
Additionally we show that in high-quality films of overdoped La$_{2-x}$Sr$_{x}$CuO$_{4}$, oxygen vacancies strongly deform the oxygen environment around the copper ions.
This observation is consistent with previous defect-chemical studies, and calls for further investigation of the defect induced properties in the overdoped regime of the hole-doped lanthanum cuprates.

\end{abstract}

\maketitle


One of the greatest challenges in research on high-temperature superconductivity is separating the effects of chemical disorder and electronic correlations on the physical properties of the superconducting cuprates.
All cuprates share two substructures \cite{keimer2015,orenstein2000}: the copper oxide (\cuo) planes, which host the valence electron system, and the so-called charge reservoir, whose chemical composition is modulated by introducing impurities such as heterovalent cations, excess oxygens, or oxygen vacancies.
The \cuo\ layers are insulating and antiferromagnetic if the Cu ions are in the valence state Cu$^{2+}$.
The imbalance between the valence of dopant and host ions in the charge reservoir results in extra charges which are transferred into the neighboring \cuo\ sheets via apical oxygens. 
As a result, the \cuo\ planes are populated with carriers and become metallic and superconducting while preserving shorter antiferromagnetic correlations. 

Among the superconducting copper oxides La$_{2-x}$Sr$_{x}$CuO$_{4}$ (LSCO) is unique 
due to the fact that the carrier concentration can be finely tuned by varying the Sr concentration, unlike other cuprates with higher T$_{c}$ that are doped by oxygen off-stoichiometry. 
A solid solution can be formed over a wide range of Sr concentration 0$<x<$1.3 \cite{sato2000}. 
This allows one to cover the full phase diagram across the whole superconducting dome from the undoped parent compound La$_{2}$CuO$_{4}$ to the highly overdoped regime where eventually superconductivity is lost. 

Recent work on LSCO thin films has revealed an unusual temperature dependence of the superfluid density in highly overdoped LSCO \cite{bozovic2016}, triggering a controversial discussion about the role of disorder in this regime of doping \cite{lee-hone2017,bozovic2016}.
In this context, it is important to realize that Sr doping modifies not only the carrier concentration but also the defect concentration \cite{smidskjaer1987,maier1991}.
Indeed from a defect chemistry point of view, LSCO may also contain oxygen vacancies (\ov) according to the equation \cite{maier1991}:
\begin{equation}
\mathrm{2SrO + 2La_{La}^{\times} + O_{O}^{\times} \longrightarrow La_{2}O_{3} + 2Sr'_{La} + V_{O}^{\bullet\bullet}}
\end{equation}
Here $\mathrm{La_{La}^{\times}}$ and $\mathrm{O_{O}^{\times}}$ are neutral La and O atoms at La and O sites respectively.
According to equation (1), the two Sr atoms at the La site (charged with -1: $\mathrm{Sr'_{La}}$) may generate an oxygen vacancy (charged with +2: $\mathrm{V_{O}^{\bullet\bullet}}$) instead of two holes.
The number of oxygen vacancies in bulk and thin films ($\delta$) can be reduced by means of annealing under high-pressure oxygen or during epitaxial growth under a reactive oxidant atmosphere such as ozone or oxygen plasma \cite{torrance1988,sato2000,bilbro2011}.
These treatments, as it has been reported, extend the superconducting dome to higher doping levels both in bulk crystals and epitaxial thin films. 
The crystal structure of LSCO without interstitial oxygen (Figure 1(a)) shows that there are two different oxygen positions: (i) apical oxygen O(2) and (ii) basal oxygen O(1).
The role of these two inequivalent oxygens has been the subject of many studies aimed at understanding the influence of these two oxygens on the physical and chemical properties such as superconductivity and oxygen vacancy formation, respectively \cite{weber2010,pavarini2001,sakakibara,peng2016}.

During the last decade it was demonstrated that high quality LSCO epitaxial thin films and heterostructures can be grown by using ozone-assisted atomic layer-by-layer molecular beam epitaxy (ALL-MBE) \cite{bozovic2016,baiutti2015}.
Samples are generally fully oxidized over a wide range of doping, since the films are grown under a high purity ozone atmosphere that even results in excessive interstitial oxygen staging in underdoped LSCO \cite{sato2000}.
Typically the concentration of carriers (holes) $p$ in these films has been estimated as the concentration of Sr dopants, $x$ ($p=x$), even in the overdoped regime \cite{bozovic2016}.
However, the oxygen deficiency induced by Sr doping in LSCO has not been widely studied, and is thus not well-understood. One practical difficulty is that the presence of oxygen deficiency is difficult to detect. 
\begin{figure*}[t!]
	\includegraphics[width=5 in]{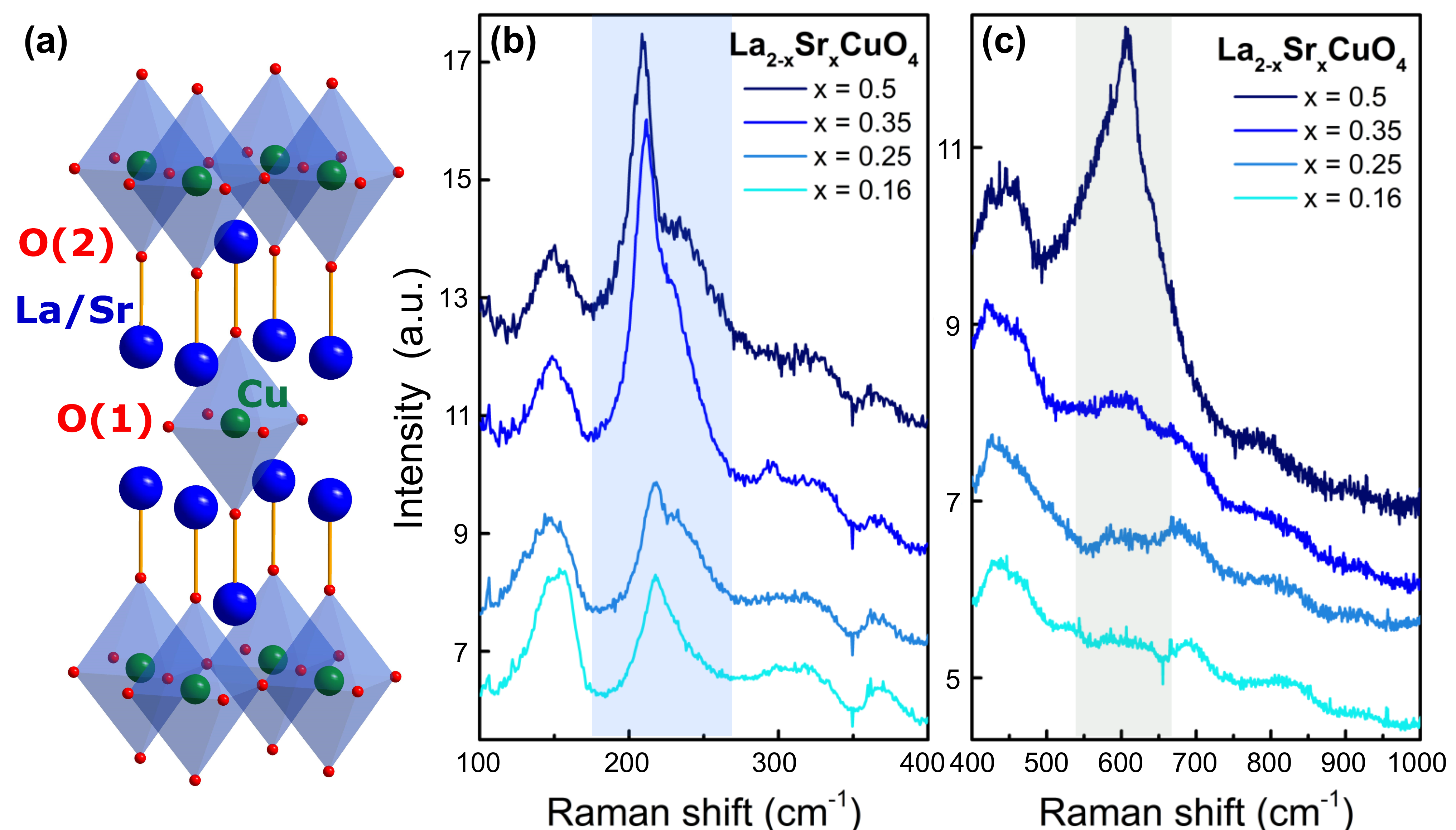}
	\caption{
		(a) The lattice structure for LSCO with tetragonal symmetry. 
		(b), (c) Unpolarized Raman spectra from LSCO thin films with different doping levels. 
		Spectra from x = 0.25, 0.35, 0.5 are shifted in the $y$-direction with 1, 1.4, 2.8 units for better visibility.
	}
\end{figure*}

Here we show that it is possible to selectively change the oxygen vacancy concentration and control it via \textit{ex-situ} postannealing. We also demonstrate that the $p=x$ assumption has a progressively more limited validity upon moving towards the overdoped regime.
This information could be essential for further understanding the physics and chemistry of high temperature superconducting cuprates.

Our tool of choice to study this issue is Raman scattering from LSCO thin films.
Raman spectroscopy is an extremely sensitive probe of local symmetry changes, and oxygen occupancies in thin films can be more easily controlled than in the bulk form via the annealing process\cite{lampakis2006,iliev1997}. 
We grew high quality LSCO epitaxial thin films with four different Sr doping concentrations ($x=0.16, 0.25, 0.35$ and $0.5$) on LaSrAlO$_{4}$ (001) single crystalline substrates (Crystal GmbH) by using the ozone-assisted ALL-MBE system (DCA Instruments) \cite{baiutti2015}.
Each film is 100 unit cells thick and all growths were controlled by using \textit{in-situ} reflection high energy electron diffraction (RHEED).
The film quality was confirmed by using atomic force microscopy (AFM) and high resolution X-ray diffraction (XRD).
During growth the substrate temperature was kept at 630 \degC\ according to the radiative pyrometer and the pressure was $\sim 1\times10^{-5}$ Torr. 
A series of post-annealing treatments was carried out in the growth chamber of the MBE system: the base pressure was $\sim 1\times10^{-8}$ Torr for each vacuum annealing and $\sim 1\times10^{-5}$ Torr for annealing in ozone.
During each growth and ozone annealing run, ozone was  
supplied from a dedicated delivery system by evaporating liquid ozone.
The diamagnetic response and resistance were measured simultaneously as a function of temperature in the range of 4.2-300 K using a motorized dip-stick.
The temperature was varied by inserting the dip-stick into the transport helium dewar.
The Raman spectra were measured with a Jobin-Yvon LabRam HR800 spectrometer (Horiba Co.) combined with a dedicated confocal microscope with motorized objective lens with short depth of focus that allows measurements of films with thicknesses of $\sim$\ 10 nm \cite{hepting2014,hepting2015}.
The samples were illuminated with a He-Ne laser with wavelength 632.8 nm, and the scattered light was collected from the sample surface with a 100x objective.
The experiments were performed in backscattering geometry along the crystallographic c-axis. 



\begin{figure*}[t!]
	\includegraphics[width=5 in]{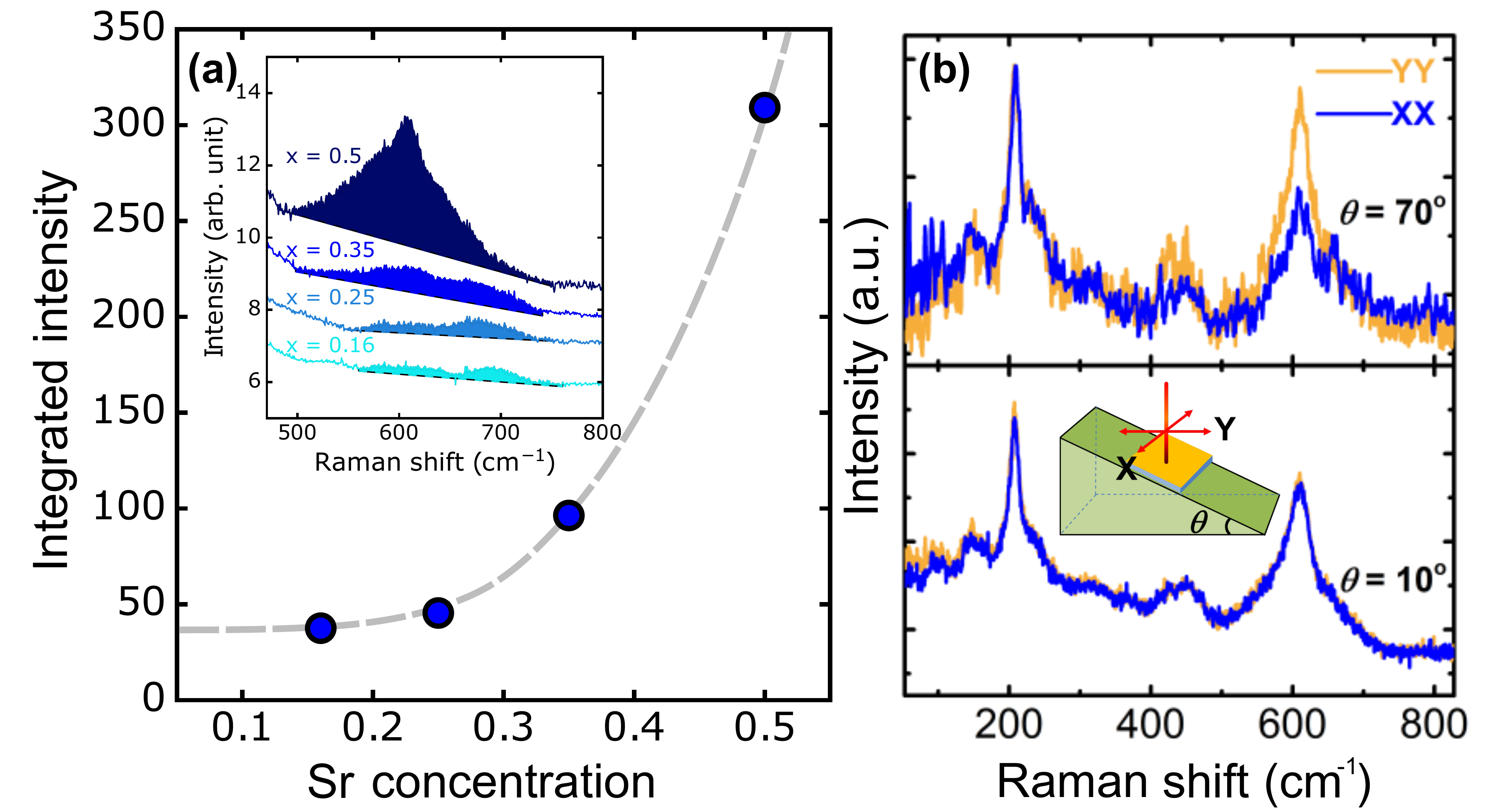}
	\caption{Analysis on the peak at $\sim$\ 600 \wn.
		(a) Integrated intensity as a function of Sr concentration.
		The gray dashed line is a guide to the eye.
		(b) Comparison between spectra from 10 $\!^\circ$ and 70 $\!^\circ$. the spectra are scaled to match the intensity of the peak at 220 \wn.
		(inset) Experimental setup for the angle dependence measurement. Yellow surface is the thin film sample.
	}
\end{figure*}

Fig. 1 shows unpolarized Raman spectra of our LSCO films with Sr concentrations $x=0.16, 0.25, 0.35$ and $0.5$.
The spectra are vertically shifted for clarity.
All Raman spectra have overall similar shapes in spite of the wide variation of $x$, and are consistent with previous reports on bulk samples \cite{sugai1989,weber1988,brun1987}.
The strongest doping dependence is observed for the mode at $\sim$ 220 \wn and the mode at $\sim$ 600 \wn: both peaks increase in intensity as the Sr concentration increases. 
The other phonons at $\sim$ 150 \wn, $\sim$ 320 \wn, $\sim$ 360 \wn, and $\sim$ 450 \wn\ are only weakly doping dependent.
The peak at $\sim$ 220 \wn\ has been assigned as the $A_{1g}$ phonon mode that involves the even motion of the La/Sr atoms \cite{sugai1989,weber1988,brun1987} bonded to the apical oxygens O(2).
Thus the change of the amplitude of this peak with Sr doping level can be interpreted either as a consequence of increasing Sr concentration or the formation of oxygen vacancies in the apical position.
The amplitude change is even more dramatic for the high energy mode at $\sim$ 600 \wn\ presented in Figure 1(c).
This broad peak is present over the whole range of $x$ and its spectral weight increases with $x$.
This particular mode has not been systematically studied in LSCO and its origin and assignment remain controversial.
In Refs.\cite{sugai2003,sugai1989} an analogous high energy mode was observed in overdoped LSCO samples and assigned to oxygen vibrations, because the high energy of the mode suggests a small atomic mass.

Interestingly, a similar broad peak located in the same energy range was observed in a number of Raman studies on different materials with similar crystal structure, such as LaSrAlO$_{4}$ \cite{hedjiev1997} and electron-doped cuprates with $T'$ structure \cite{heyen1991}.
Although the origin of this peak has not been conclusively established, previous studies suggest a connection with the presence of defects for this phonon mode with \aog\ symmetry and they point out that it has a larger contribution with light polarized along the z-axis.
Here we use the same notation for this phonon mode as \asg\ mode following ref. \cite{heyen1991}. 
The evolution of the peak at $\sim$ 600 \wn\ with Sr concentration possibly supports the hypothesis of a defect-induced origin of the \aog\ mode.
As this peak partially overlaps with the peak at 700 \wn\, which is essentially constant over the whole doping range, we plot the integrated area of the \asg\ peak as a function of Sr concentration in Figure 2 (a).
Taking into account the defect chemistry analysis on the LSCO compound  \cite{maier1991}\, the enhancement of spectral weight is consistent with the presence of more oxygen vacancies.
In order to further clarify the symmetry and origin of the phonon mode at $\sim$ 600 \wn\ in LSCO thin films, Raman spectra with $zz$-polarization should be measured. 
Unfortunately the planar sample geometry of thin films makes it extremely challenging to collect the Raman signal from the out of plane $z$ component.

We have overcome this challenge by measuring Raman spectra on tilted samples as described in Figure 2(b).
We positioned LSCO ($x=0.35$) on wedges with different angles $\theta$.
In this configuration we could pick up a partial contribution from the $z$ direction that increases with the angular deviation from the original standard z(YY)$\underbar{z}$ geometry (in Porto's notation).
The result of the angular dependent measurements is presented Figure 2(c), where a striking difference between XX and YY polarized spectra is clearly seen for $\theta$ = 70 $\!^\circ$ at $\sim$ 600 \wn.
The phonon mode at $\sim$ 600 \wn\ has a larger $z$ component than the \aog\ phonon mode at $\sim$ 220 \wn.
This observation is consistent with the above-cited previous studies on similar compounds. Along the lines of Ref. \cite{heyen1991} we can assign this anisotropic phonon mode \asg\ to the vibration of the apical oxygens at the vertices of the octahedral oxygen network surrounding the copper ions.
Hence the spectral weights of \aog\ and \asg\ phonon modes in Raman spectra can be used as selective and sensitive markers to judge the local modifications of the crystal structure caused by formation of oxygen vacancies in the LSCO epitaxial thin films.
In the following, we focus on these markers to evaluate the tunability of oxygen concentration via low temperature annealing under vacuum.

\begin{table*}[t]
	\caption{\label{tab:table1}%
		The summary of mutual inductance measurements and XRD of LSCO thin films. The \tc$_{,middle}$ is defined as the temperature where the imaginary part of the mutual inductance is maximized. This \tc$_{,middle}$ is consistent with previous reports on the phase diagram of LSCO thin films grown by oxide MBE.
	}
	\begin{ruledtabular}
		\begin{tabular}{ccccc}
			Nominal doping&Treatment &T$_{c,onset}\mathrm{(K)}$&T$_{c,middle}\mathrm{(K)}$&$c$(\AA)\\ \hline
			&as grown&40&37&13.277\\
			$x$=0.16&1st annealed at 280 \degC\ for 30 mins in vacuum&40&34&-\\
			&2nd annealed at 280 \degC\ for 30 mins in vacuum&26&$<$ 4.2&13.259\\
			&ozone annealed  at 630 \degC\ for 1 hrs in Ozone&39&34&13.289\\ \hline
			&as grown&31&25&13.24\\
			$x$=0.25&1st annealed at 280 \degC\ for 20 mins in vacuum&30&24&13.24\\
			&2nd annealed at 280 \degC\ for 40 mins in vacuum&28&21&13.252\\ 
			&ozone annealed  at 630 \degC\ for 1 hrs in Ozone&30&25&13.297\\\hline
			&as grown&15&10&13.29\\
			$x$=0.35&1st annealed at 280 \degC\ for 30 mins in vacuum&17&13&13.277\\
			&2nd annealed at 280 \degC\ for 240 mins in vacuum&$<$ 4.2&$<$ 4.2&13.265\\
			&ozone annealed  at 630 \degC\ for 1 hrs in Ozone&15&10&13.285\\
		\end{tabular}
	\end{ruledtabular}
\end{table*}

Using a series of vacuum annealing processes we could vary the concentration of oxygen vacancies in our films.
After each annealing process the superconducting transition temperature \tc, the diamagnetic response and XRD were measured.
In Table 1 we summarize samples, details of the annealing process, superconducting transition temperatures and $c$-axis lattice constants as obtained from XRD measurements.
We studied three LSCO samples with different Sr doping levels $x=0.16$ (optimum doping), $0.25$ and $0.35$ (overdoped).
In Tab. 1 two superconducting transition temperatures are listed: \tc$_{,onset}$ is the temperature at which the real and imaginary parts of mutual inductance start to change due to the diamagnetic screening, whereas \tc$_{,middle}$ corresponds to the temperature where the imaginary part of the mutual inductance is maximized. 
Vacuum annealing processes reduce \tc\ in all samples.
Only in the case of overdoped LSCO with $x=0.35$ the first vacuum annealing slightly increased \tc\ due to the compensation of extra holes by electrons from \ov.
Nonetheless, a second vacuum annealing process suppressed \tc\ underreaching the \tc\ of the optimum doped sample.
This already indicates that the lattice distortions induced by \ov\ play an important role in determining \tc\ together with the carrier concentration in the \cuo\ planes.
\tc\ was recovered by annealing in ozone as shown in Table 1 for the optimum doped LSCO film.
The reversibility of \tc\ suggests that the formation of oxygen vacancies in high quality LSCO films is also reversible.
Notably, while \tc\ was suppressed remarkably after a second annealing in vacuum, the reduction of the $c$-lattice constant was only moderate.

In order to clearly demonstrate that these peaks stem from oxygen deficiency, we looked at the Raman modes $\sim$ 220 \wn, and $\sim$ 600 \wn\ after each vacuum annealing that can modify the concentration of oxygen ions in the thin films.
The un-polarized Raman spectra for three LSCO films $x=0.16, 0.25$ and $0.35$ measured after consecutive annealing processes are presented in Figure 3 (a), (b), (c) respectively.
For clarity the spectra are vertically shifted.
For comparison we plot also the spectrum obtained from the bare LaSrAlO$_{4}$ substrate (gray line) that contributes to the overall Raman signal from the samples.
As mentioned above the strongest doping dependence of the spectral weights in the as grown LSCO films was observed at $\sim$ 220 \wn\ (blue-shaded area), and $\sim$ 600 \wn\ (gray-shaded).
Subsequent annealing of LSCO films in vacuum at 280 \degC\ has a dramatic effect on both modes in all three samples.

The spectral weight of the mode at $\sim$ 220 \wn\ is progressively suppressed with each vacuum annealing due to the increase in \ov\ at the apical site that, in turn, affects the motion of La/Sr atoms.
Annealing in ozone recovers this peak to the initial value for all three doping concentration, as shown in Figure 3.
On the other hand the spectral weight of the \asg\ phonon mode at $\sim$ 600 \wn\ gradually increases after the vacuum annealing processes,
differently from the in-plane modes at $\sim$ 320 \wn\ and $\sim$ 370 \wn\ which do not vary with doping and thermal treatment. We can thus conclude that our findings imply a selective generation of oxygen vacancies at the apex of the octahedron, O(2) in the Figure 1(a).
This is supported by the changes of the apical-related modes at $\sim$ 220 and $\sim$ 600 \wn\ and by the fact that the other planar modes remain unchanged.

In conclusion, our study has provided new insights into the defect chemistry of high quality LSCO epitaxial films grown by ozone-assisted MBE: (i) the number of oxygen vacancies \ov\ increases strongly as a function of Sr doping even in samples grown in a highly oxidizing environment; (ii) the oxygen vacancies can be selectively introduced at the apical position of the CuO$_{6}$ octahedron by low temperature annealing in vacuum; (iii) confocal Raman microscopy is an appropriate tool to investigate oxygen defects in thin epitaxial oxide films. 
\begin{figure*}[!htb]
	\includegraphics[width=5 in]{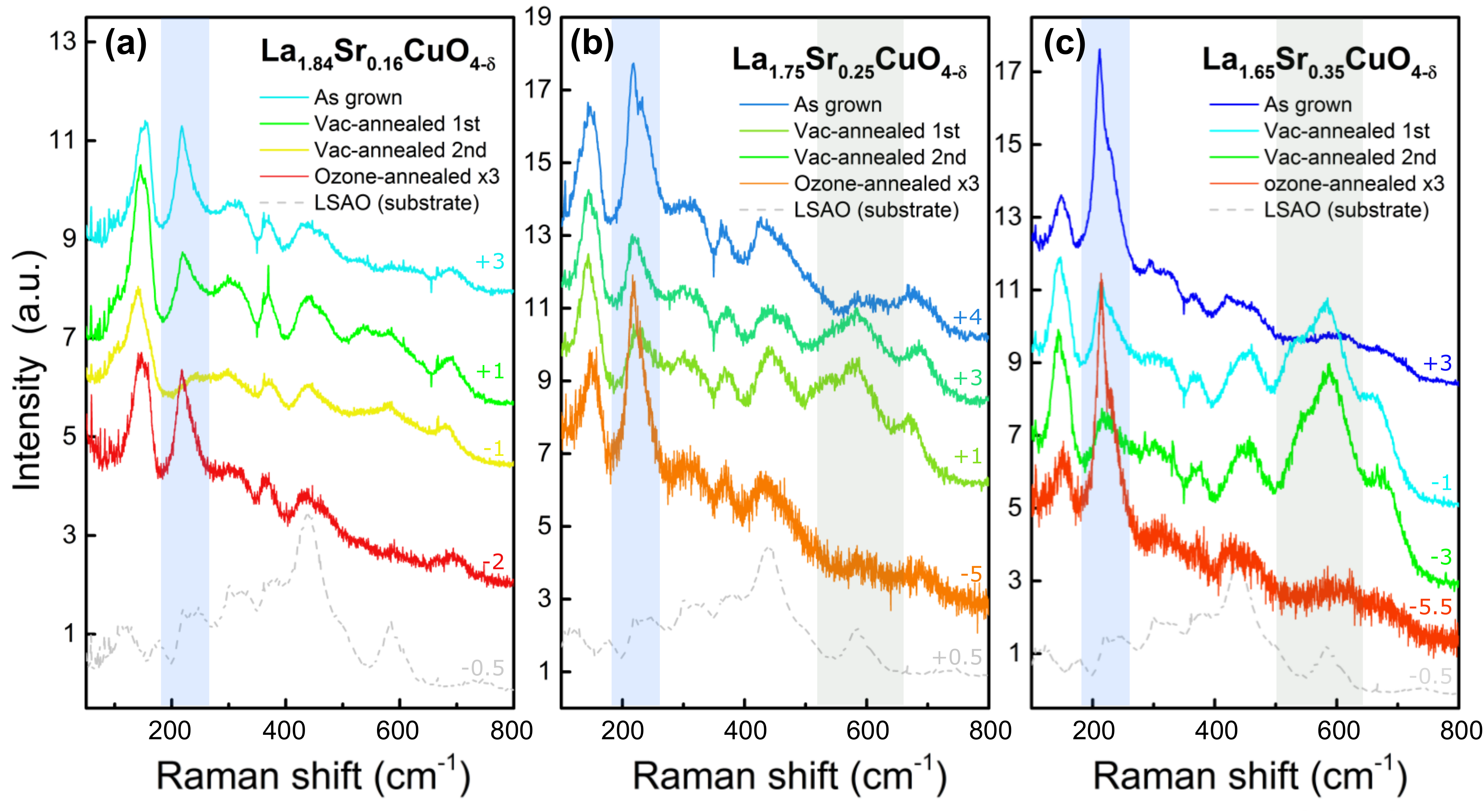}
	\caption{Raman spectra from annealed samples with different Sr concentration
		(a) x=0.16.
		(b) x=0.25
		(c) x=0.35.
		Spectra are shifted in $y$-direction for better visibility. The numbers above spectra are the amounts of shifts in arbitrary units.
	}
\end{figure*}

Moreover the increasing tendency to form \ov\ in high quality LSCO thin films as $x$ increases highlights an important source of disorder that one needs to consider in interpreting the physical properties of transition metal oxides.
We used the prototypical layered transition metal oxide LSCO to investigate the effect of annealing in different oxygen environments on the defect concentration. 
The higher tendency of overdoped samples to accomodate \ov\ calls for further investigations of the influence of disorder on the superconducting properties in this regime of the phase diagram \cite{bozovic2016,lee-hone2017}. 
Finally the selective generation of apical oxygen vacancies could provide an \textit{in situ} tool to tune the electronic structure of LSCO without major disruption of the \cuo\ plane \cite{wei2016}. 
 



\begin{acknowledgments}
We thank P. Specht, B. Stuhlhofer for technical support, and J. Sauceda for fruitful discussions.  
\end{acknowledgments}

\bibliography{LSCO_Raman}

\end{document}